\documentclass[twocolumn,showpacs,preprintnumbers,amsmath,amssymb]{revtex4}
\usepackage[english]{babel}
\usepackage[latin2]{inputenc}
\usepackage{epsfig,color}
\begin{document}
\preprint{Czech. J. Phys. {\bf 53} (2003) 33-39.}
\title{The influence of hydrogen adsorption\\ on magnetic
properties of Ni/Cu(001) surface}
\author{Franti\v{s}ek
M\'aca, Alexander B. Shick} \affiliation{Institute of Physics
ASCR, Na Slovance 2, CZ-182 21 Praha 8, Czech Republic}
\author{Josef Redinger, Raimund Podloucky and Peter Weinberger}
\affiliation{Center for Computational Materials Science, TU
Vienna, Getreidemark 9, A-1060 Vienna, Austria}


\begin{abstract}
Ni/Cu(001) is known as a unique system showing the
spin-reorientation transition from an in-plane to out-of-plane
magnetization direction when the Ni-overlayer thickness is
increased. We investigate different relaxed multilayer structures
with a hydrogen adlayer using the full-potential linearized
augmented plane-wave method. The relaxed geometries, determined by
total energy and atomic force calculations, show that H-monolayer
strongly influences the interlayer distance between the Ni-surface
and sub-surface layers yielding the outward relaxation of Ni-layer
at H/Ni interface. Furthermore, large decrease of local magnetic
moments at the top surface area is found for the surface covered
by H. The magneto-crystalline anisotropy energies calculated for
fully relaxed H/Ni-films. The spin-reorientation transition
critical thickness of 4 ML is found in good quantitative agreement
with the experiment.
\end{abstract}
\maketitle

\section{Introduction}     

Ultrathin ferromagnetic films grown on nonmagnetic substrate show
peculiar magnetic behavior \cite{bluegel}. One of the unique
phenomena which is observed in the ultrathin Ni films on Cu(001)
substrate is the spin direction reorientation transition (SRT)
from an in-plane to out-of-plane magnetization direction when the
Ni-overlayer thickness is increased \cite{babe, kirs99, kirs00}.
It is very important for the spintronic magnetic device
applications and its microscopic understanding attracted recently
both experimental and theoretical \cite{uibe99, shic99} interest.
The gas adsorption on magnetic film provides the way to monitor
the surface magnetic properties resulting in strong influence on
the SRT critical thickness $d_c$. For Ni/Cu(001) films, the
hydrogen adsorption is observed to reduce $d_c$ by $\approx$ 4
monolayers (ML) from its value of $\approx$ 11 ML for the
vacuum/Ni/Cu-films \cite{kirs99, kirs00}.

The key quantity which drives the SRT in ultrathin magnetic films
is the magneto-crystalline anisotropy energy (MAE). It determines
the preferred film magnetization orientation by minimizing the
free energy of the system. For the free-standing Ni/Cu(001) films,
the $d_c$ is found to be determined \cite{uibe99, shic99} by
competition of the uniaxial MAE due to the tetragonal distortion
of Ni-films (so called ``volume" MAE) and the surface MAE. The
adsorption of a gas produces non-trivial changes in geometrical
structure as well as magnetic properties of Ni films \cite{wimm85,
wein85} influencing both volume and surface MAE, and results in
change of the SRT critical thickness.

In recent years it has become possible to make use of {\em ab
initio} density functional theory to predict the MAE in ultra-thin
films \cite{jansen}. From theoretical and computational point of
view, the MAE calculations are extremely difficult due to the very
high energy resolution in the range of few $\mu $eV which is
required. For the H/Ni/Cu(001) films, the problem becomes even
more complex due to the need of accurate account of structural
relaxation. It requires us to use the most accurate total energy
full-potential linearized augmented plane wave method (FP-LAPW)
\cite{wimm85} in order to account simultaneously on equilibrium
geometrical and magnetic structures.

\section{Method and Results of Calculations}

The experiments show that H$_2$ adsorbs dissociatively on Ni(001)
surface in fourfold hollow sites \cite{sten85}. Therefore, as a
structural model for {\it ab-initio} calculations we use
free-standing Ni-films (H/Ni$_d$/H, $d$ = 1-11) and Ni-films on
Cu-substrate (H/Ni$_d$/Cu$_7$/Ni$_d$/H, $d$ = 1--6) with ordered
$p(1\times 1)$H adsorbate overlayer as shown in Fig. 1.
%
%
\begin{figure}[h]                     
\begin{center}                         
 \epsfig{file=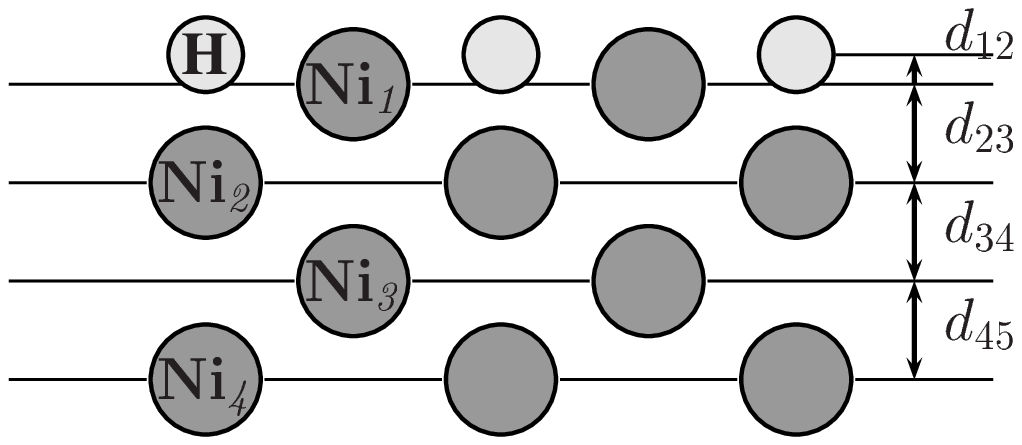,height=2.81cm,width=6.cm,angle=0}
\end{center}                         
\vspace{-2mm} \caption{Surface layers of H/Ni$_n$/Cu(100) system.}
\end{figure}                        

 The in-plane
experimental lattice constant of Cu $a_{Cu} = 3.615$ \AA
~\cite{LB} is used which remains unchanged in the calculations.
All interlayer spacings $d_{ij}$ (see Fig. 1) are relaxed to the
equilibrium values. In order to determine the equilibrium slab
geometry we employ the FP-LAPW method in FLAIR implementation
(unpublished improved and rewritten version of the original FLAPW
codes \cite{wimm85}). The scalar-relativistic atomic force
technique is employed for the total energy minimization. Here, the
Perdew and Wang \cite{pw91} approximation for exchange-correlation
potential is used with the plane wave energy cutoff $E_{cut}$ of
13 Ry, and the 28 special k-points in the 1/8th irreducible part
of 2-dimensional Brillouin zone (2D BZ) are used for the BZ
integrations. The convergence better than 1 $\times$ 10$^{-6}$
e/(a.u.)$^3$ is achieved for charge/spin densities, and better
than 0.2 mRy/a.u. for the atomic force acting on individual atom.

The relaxed interlayer distances for different free-standing films
(H/Ni$_n$/Cu$_7$/Ni$_n$/H, $n$ = 3--6) are shown in Fig. 2 and
compared with the relaxation of tetragonal $p(1\times 1)$H Ni(100)
surface. Clearly, we can distinguish the outward relaxation of top
Ni layer as well as the strong influence of Ni/Cu interface on
inter-layer distances in the close vicinity of the interface. The
relaxation of Ni/Cu subsurface is different only for 5 ML Ni. The
deeper interlayer distances (thicker film) approach the bulk
value.
%
%
\begin{figure}[t]                     
\begin{center}                         
\epsfig{file=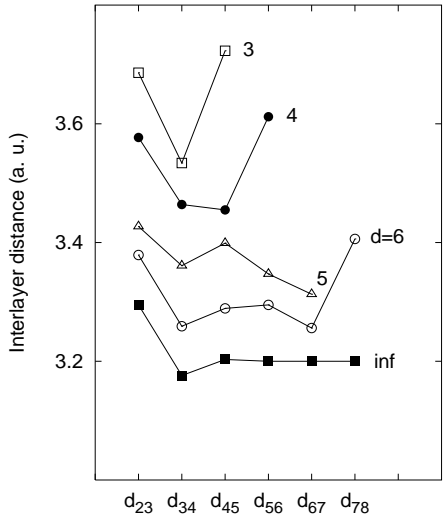,height=7.5cm,width=6.cm,angle=0}
\end{center}                         
\vspace{-2mm} \caption{Relaxation of interlayer distances for
H/Ni$_n$/Cu(001) for $n=3-6$ and for $n=\infty $. For clarity, the
data sets are shifted by 0.1 a.u. Lines serve as guide to the
eye.}
\end{figure}                        
%
%
\begin{table}[ht]                          
\caption{Interlayer distances in a.u.} \vspace{2mm}
\small
\begin{center}                             
\begin{tabular}{|c|c|c|}           
 \hline ~&~&~\\[-2mm]
~&$\bf HNi_{11}H$&$\bf HNi_6Cu_7Ni_6H$\\[1mm] \hline
$d_{12}$&0.63&0.60\\ \hline $d_{23}$&3.30&3.28\\ \hline
$d_{34}$&3.18&3.16\\ \hline $d_{45}$&3.20&3.19\\ \hline
\end{tabular}                            
\vspace{-1mm}
\end{center}                             
\end{table}

 The equilibrium values of inter-layer distances for two model
systems: eleven-layer tetragonal Ni-film covered with H and
seven-layer Cu-film with overlayer of six Ni layers covered with H
are shown in Table 1.                            
Both interfaces (H/Ni, Ni/Cu) influence strongly the interlayer
distances in the overlayer slab. Note, that the calculated bond
length $d_{H-Ni}$ of 3.48 a.u. is slightly shorter than the bond
length for the bulk nickel hydride (3.52 a.u.). The outward
relaxation of top Ni-layer is found, the H-ML influences strongly
only the interlayer distance between the Ni-surface and
Ni-subsurface layer. Below the interface, the interlayer distances
oscillate around the bulk value for strained tetragonal Ni bulk
($d_\perp = 3.20$ a.u.). The electron screening in the metal is
responsible for fast damping of these oscillation.

%
%
\begin{table}[h]                          
\caption{Magnetic moments in $\rm \bf \mu_B $} \vspace{2mm}
\small
\begin{center}                             
\begin{tabular}{|c|c|c|}           
 \hline ~&~&~\\[-2mm] ~&$\bf HNi_{11}H$&$\bf HNi_6Cu_7Ni_6H$\\[1mm] \hline
$\bf Ni_{\it 1}$&0.238&0.239\\ \hline $\bf Ni_{\it
2}$&0.593&0.588\\ \hline $\bf Ni_{\it 3}$&0.640&0.636\\ \hline
$\bf Ni_{\it 4}$&0.633&0.633\\ \hline
\end{tabular}                            
\vspace{-1mm}
\end{center}                             
\end{table}                             
The layer-resolved spin magnetic moments $M_s$ for these systems
are shown in Table 2. These values correspond to the magnetic
moments in the Ni ``muffin-tin" spheres ($R_{MT}=2.2$~a.u.).
 It is
seen that there is strong reduction of the spin magnetization for
the top Ni-layer due to the interaction with the H-adlayer. The
strong hybridization of the H {\em s} state with the filled
majority Ni {\em d} band changes the band structure and the
surface density of states (SDOS) leading to the decrease of
spin-majority and increase of the spin-minority SDOS. Away from
the interface, the spin magnetic moments are slowly converging to
their bulk values. The layer-resolved magnetic moments for
different H/Ni$_n$/Cu$_7$-films ($n$ = 3--6) on Cu-substrate are
shown in Fig. 3. We note that for very thin H/Ni$_n$/Cu$_7$-films
with n=1,2 the Ni-local moments disappear and the system becomes
non-magnetic. With the increase of the Ni-film thickness, the
local Ni $M_s$ of 0.24 $\mu_B$ is formed at the Ni-interface and
then increases away from the H/Ni interface. When approaching the
Ni/Cu interface, the Ni-atom magnetic moments start to decrease
again and become $\approx$ \mbox{0.45 $\mu_B$} for Ni/Cu interface
layer (depending of the Ni-film thickness).
%
%
\begin{figure}[t]                     
\begin{center}                         
 \epsfig{file=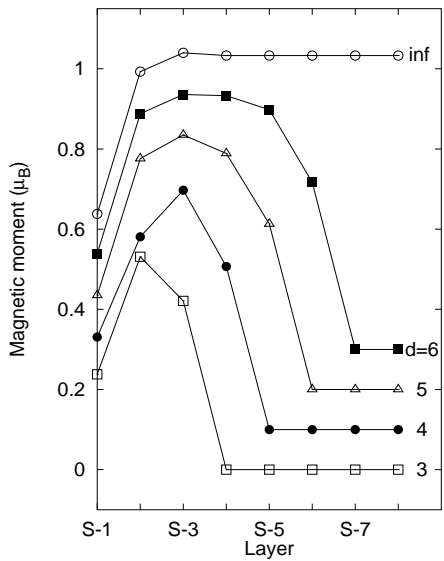,height=7.5cm,width=6.cm,angle=0}
\end{center}                         
\vspace{-2mm} \caption{Layer-resolved magnetic moments of
H/Ni$_n$/Cu(001) for $n=3-6$ and for $n=\infty $. For clarity, the
data sets are shifted by $0.1(n-1)\mu _B$.Layers are labeled S,
S-1, \dots ~, starting with the surface layer S. Lines serve as
guide to the eye.}
\end{figure}                        

The anisotropic energy density of a tetragonal ferromagnetic film
is written as \cite{Landau}:

\begin{eqnarray*}
 E/V \; = - K_1^v m_z^2 \; - \; K_2^v m_z^4 \; -
\; K_3^v m_x^2 m_y^2 \;\\ - \frac{2}{d} (K_1^s m_z^2 \; + \; K_2^s
m_z^4 \; + \; K_3^s m_x^2 m_y^2) ,\end{eqnarray*}

 where, $K^v$ terms are 2nd
and 4th-order volume-type anisotropy constants, $K^s$ terms are
surface/interface-type 2nd and 4th-order anisotropy constants,
$m_{x,y,z}$ are magnetization cosines with respect to the crystal
axes, and $d$ is a thickness of magnetic film. We assume that the
4th-order terms in Eq.(1) are significantly smaller than 2nd-order
uniaxial anisotropy constant $K_1^v$, and neglect them. The MAE
then can be characterized with the difference in the total energy
when magnetization is oriented along [100] (in-plane, $\parallel$)
and [001] (out-of-plane, $\perp$) axes (MAE = $E[100] \; - \;
E[001]$).

We use the relativistic version \cite{flapwso} of the FP-LAPW
method to solve self-consistently the Kohn-Sham-Dirac equations
with spin-orbit coupling included to obtain the ground state
charge and spin densities for the magnetization directed along
[001]-axis. The fully optimized interlayer distances as discussed
above are used for free-standing Ni-films and Ni-films on
Cu-substrate with H-adlayer. The MAE is obtained by applying the
force theorem to the spin - axis rotation \cite{Igor}: from the
self-consistent ground state charge and spin density obtained for
the [001] spin axis, a calculation of the band structure for [100]
spin axis orientation is performed, and difference of the single
particle eigenvalue sums is then taken to be the MAE. For the MAE
calculations the k-points mesh equivalent to 6400 k-points in the
full 2D BZ is used guaranteeing the MAE convergence better than 10
$\mu$eV.

The MAE as a function of the Ni-film thickness for the H/Ni$_d$/H
films with $d$ = 3,5,7,9 ML is shown in Fig. 4. For the $d=3$ we
found small and negative MAE \mbox{(-0.02 meV)} keeping the
Ni-film magnetization in [100] plane. It becomes positive for the
$d=5$ (0.432 meV) resulting in the out-of-plane magnetization
switching. The MAE is positive with further increase of $d$ and
shows pronounced oscillations. The linear interpolation MAE = $ d
\cdot K_V \; + \; 2 \cdot K_I$ yields the estimates for the
``volume" $K_V$= 0.106 meV/atom and ``interface" $K_I$=-0.127
meV/atom MAE contributions. The calculated ``volume" MAE agrees
well with $K_V$= 0.0835 meV/atom calculated for free-standing
unrelaxed Ni-films without H-adlayer \cite{shic99} and
extrapolated to T=0 K experimental value of 0.072 meV/atom of Ref.
\cite{farle}. The H/Ni ``interface" $K_I$ is calculated to be
substantially smaller than vacuum/Ni ``surface" $K_S$ = -0.447
meV/atom \cite{shic99} for free-standing Ni-films and extrapolated
to T=0 K experimental value of -0.7 meV/atom \cite{kirs99}. We
note that for the very thin Ni-films considered here, validity of
the linear fit for the MAE and its separation into ``volume" and
``surface" contributions is {\it not well justified} due to the
strong dependence of the Ni-film magnetic properties on the
thickness of the film, especially in the presence of the
H-adlayer.
%
%
\begin{figure}[t]                     
\begin{center}                         
\epsfig{file=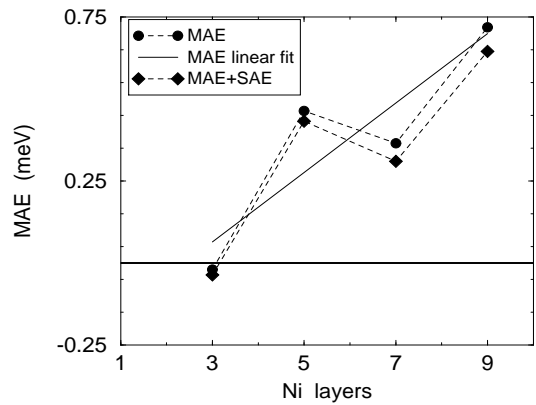,height=5.43cm,width=6.94cm,angle=0}
\end{center}                         
\vspace{-2mm} \caption{The MAE (circles) and MAE+SAE (diamonds)
for the fully relaxed H/Ni$_d$/H films as a function of the film
thickness d (in ML).}
\end{figure}                        

In order to estimate the $d_c$ critical thickness for SRT, we take
into account the shape anisotropy energy (SAE) due to the magnetic
dipole interaction, which provides additional in-plane anisotropy.
This anisotropy is estimated using the relation SAE = $-2 \pi M^2$
to the spin magnetization density M (in CGS units). It yields for
the SAE the values of -0.016 meV, -0.032 meV, -0.055 meV and
-0.073 meV corresponding respectively to the H/Ni$_d$/H films with
$d$ = 3,5,7,9 layers (which are consistent with experimentally
derived SAE/Ni atom of -0.0075 meV \cite{kirs99}). This additional
negative SAE slightly shifts the MAE in Fig. 4 downwards yielding
the $d_c$ = 4.

Indeed, the H/Ni$_d$/H model is a way too simple to describe
quantitatively the SRT in in Ni/Cu layers with H-adlayer. As it
was already mentioned in Ref. \cite{kirs99,uibe99} the Ni/Cu
interface can play an important role. In order to evaluate the
influence of Ni/Cu interface on SRT we perform the calculations
for H/Ni$_d$/Cu$_5$/Ni$_d$/H ($d$ = 3,4) films, where the 5 Cu
layers play a role of the substrate. The use of 5 ML of Cu layers
instead of 7 ML allows to reduce computational effort without
producing any significant impact on the magnetic properties and
the MAE of H/Ni/Cu-films, since they are originated from Ni
magnetic film and not the Cu non-magnetic substrate, and the
relaxed Cu-interlayer distance below the Ni-interface is found to
be very close to its bulk value of 3.427 a.u. For the $d=3$ we
found negative MAE of -0.192 meV and SAE of -0.036 meV keeping the
Ni-film magnetization in [100] plane. Already for $d=4$ the MAE
becomes positive 0.543 meV while the SAE is small and negative
-0.052 meV. Again as in the case of free-standing H/Ni$_d$/H films
we get the SRT critical thickness $d_c=4$.

To summarize, we found that hydrogen adsorption for the Ni films
on Cu substrate yields the reduction of the SRT critical
thickness. The calculated $d_c=4$ for both H/Ni$_d$/H and
H/Ni$_d$/Cu$_5$/Ni$_d$/H films agrees well with the experimental
$d_c$ of 7 ML \cite{kirs99}. We show that the MAE has strong and
oscillatory dependence on the Ni-film thickness which deviates
substantially from the linear fit. We attribute the decrease of
the $d_c$ due to the H adsorption to originate from strong
reduction of the magnitude for the Ni ``surface" MAE contribution.
In turn, this is caused by strong decrease of the exchange
splitting at the H/Ni interface due to the strong hybridization of
the H $s$ state with the Ni-bands.

\subsection* {Acknowledgment} {\small The financial support was
provided by the Academy of Sciences of the Czech Republic (Grant
No. A1010214), CMS Vienna (GZ 45.504), and by the RTN project
"Computational Magnetoelectronics" of the European Commission
(HPRN-CT-2000-00143).}

\bigskip

\begin{thebibliography}{9}               
\bibitem{bluegel}
S. Bl\"ugel:
Phys. Rev. Lett. {\bf 68} (1992) 851 and references therein.

\bibitem{babe}
B. Schulz and K. Baberschke: Phys. Rev. B {\bf 50} (1994) 13467.

\bibitem{kirs99}
 R. Vollmer
 {\em et al.}:
 Phys. Rev. B  {\bf 60} (1999) 6277.

\bibitem{kirs00}
 S. van Dijken, R. Vollmer, B. Poelsema, J. Kirschner: J. Magn.
 Magn. Mater. {\bf 210} (2000) 316.

\bibitem{uibe99}
 C. Uiberacker
 {\em et al.}:
  Phys. Rev. Lett. {\bf 82} (1999) 1289.

\bibitem{shic99}
 A.B. Shick, Y.N. Gornostyrev, A.J. Freeman: Phys. Rev. B  {\bf 60} (1999) 3029.
\bibitem{wimm85}
E. Wimmer, H. Krakauer, A.J. Freeman: Adv. Electron. Electron
Phys. {\bf 65} (1985) 357.
 \bibitem{wein85}
 M. Weinert and J.W. Davenport: Phys. Rev. Lett. {\bf 54} (1985)
 1547.
\bibitem{jansen} H.J.F. Jansen: Phys. Rev. B {\bf 59} (1999) 4699
and references therein.
\bibitem{sten85}
I. Stensgaard and F. Jakobsen: Phys. Rev. Lett. {\bf 54} (1985)
711.
 \bibitem{LB}
 W.B. Pearson: {\it A handbook of
lattice spacings and structures of metals and alloys}. Pergamon
Press, Oxford, 1958.
 \bibitem{pw91}
 J.P. Perdew and Y. Wang: Phys. Rev. B {\bf 45} (1992) 13244.

\bibitem{Landau}
 {L. D. Landau, E. M. Lifshitz, and L. P. Pitaevskii:}
  {\it Electrodynamics of Continuous Media}.
  Oxford, 1995, p.138.
\bibitem{flapwso}
A. B. Shick, D. L. Novikov and A. J. Freeman: Phys. Rev. {\bf B
56} (1997) R14259.

\bibitem{Igor} I. V. Solovyev, P. H. Dederichs and I. Mertig:
Phys. Rev. {\bf B 52} (1995) 13419 and references therein.

\bibitem{farle} M. Farle {\it at al.}: Phys. Rev. {\bf B 55} (1997) 3708.
\end{thebibliography}
\end{document}